# Automating the Training and Deployment of Models in MLOps by Integrating Systems with Machine Learning

**Penghao Liang[1*], Bo Song[1,2], Xiaoan Zhan[1,3], Zhou Chen[2], Jiaqiang Yuan[3]**

[1*] Information Systems,Northeastern University,San Jose, CA ,USA
[1.2] Computer Science,Northeastern University, Boston, MA,USA
[1.3] Electrical Engineering, New York University, NY, USA
[2] Software Engineering,Zheiiang University,Hangzhou.China
[3] Information Studies, Trine University, Phoenix, AZ,USA

*Corresponding author:Penghao Liang,-mail:liang.p@northeastern.edu

**Abstract.**

This article introduces the importance of machine learning in real-world applications and explores the rise of MLOps (Machine Learning Operations) and its importance for solving challenges such as model deployment and performance monitoring. By reviewing the evolution of MLOps and its relationship to traditional software development methods, the paper proposes ways to integrate the system into machine learning to solve the problems faced by existing MLOps and improve productivity. This paper focuses on the importance of automated model training, and the method to ensure the transparency and repeatability of the training process through version control system. In addition, the challenges of integrating machine learning components into traditional CI/CD pipelines are discussed, and solutions such as versioning environments and containerization are proposed. Finally, the paper emphasizes the importance of continuous monitoring and feedback loops after model deployment to maintain model performance and reliability. Using case studies and best practices from Netflix, the article presents key strategies and lessons learned for successful implementation of MLOps practices, providing valuable references for other organizations to build and optimize their own MLOps practices.

## 1. Introduction

Machine learning has revolutionized the way people use and interact with data, driving business efficiency, fundamentally changing the advertising landscape, and revolutionizing healthcare technology. Over the past decade, machine learning (ML) has become an essential part of countless applications and services in a variety of fields. Thanks to the rapid development of machine learning, there have been profound changes in many fields, from health care to autonomous driving. However, the increasing importance of machine learning in practical applications also brings new challenges and problems, especially when it comes to moving models from a laboratory environment to a production environment. Traditional software development and operations methods often fail to meet the specific needs of machine learning models in production, resulting in challenges such as the complexity of model deployment, difficulties in performance monitoring, and the absence of continuous integration and continuous deployment [1](CI/CD) processes.

To address these issues, attention is being paid to an emerging field called Machine Learning System Operations (MLOps). MLOps is a relatively new term that has gradually gained traction over the past few years. It closely links computer systems and machine learning and considers new challenges in machine learning from the perspective of traditional systems research. [2]MLOps is not just a tool or process, it is a philosophy and methodology that aims to achieve continuous delivery and reliable operation of machine learning models. Against this background, this article will explore ways to automate model training and deployment by integrating systems with machine learning. First, we will review the challenges and problems in existing MLOps, and then lead to the topic of this article, which is how the integration of systems with machine learning can solve these challenges and improve productivity.

## 2. Related Work

*2.1. Review on the development of MLOps*

In the past, different software development process models and development methods have appeared in the field of software engineering. Prominent examples include the waterfall model and the Agile Manifesto. These approaches have a similar goal of delivering a production-ready software product. In 2008/2009, a concept called "DevOps" [3] emerged to reduce problems in software development. DevOps is not just a pure approach, but represents a paradigm for solving social and technical problems in organizations engaged in software development. It aims to close the gap between development and operations and emphasizes collaboration, communication, and knowledge sharing. It ensures automation through continuous integration, continuous delivery, and continuous deployment (CI/CD) for fast, frequent, and reliable releases. [4] In addition, it is designed to ensure continuous testing, quality assurance, continuous monitoring, logging, and feedback loops.

Due to the commercialization of DevOps, many DevOps tools continue to emerge, which can be divided into six categories: Collaboration and knowledge sharing (e.g. Slack, Trello, GitLab wiki), source code management (e.g. GitHub, GitLab), build process (e.g. Maven), continuous integration (e.g. Jenkins, GitLab) CI), deployment automation (e.g. Kubernetes, Docker), monitoring and logging (e.g. Prometheus, Logstash). Cloud environments are increasingly equipped with ready-to-use DevOps tools designed for cloud use, facilitating efficient value generation [5]. With the shift to the novel paradigm of DevOps, developers need to care about what they develop because they also need to manipulate it.

As empirical results show, DevOps ensures better software quality. People in the industry and academia have gained extensive experience in software engineering through the use of DevOps. This experience is now being used for ML automation and operations.

Therefore, the evolution of DevOps to MLOps[6] is a process of extensibility that includes the adaptation of traditional software development processes and the introduction of new technologies. In this process, organizations need to combine the principles of DevOps with the unique requirements of machine learning, redesigning and extending existing continuous integration, continuous delivery, and continuous deployment pipelines to accommodate the development, training, and deployment processes of machine learning models. [7]At the same time, specific tools and techniques for machine learning need to be introduced, such as model warehouses for model versioning, trial tracking systems for automated experiment management, and specific tools for model monitoring and logging. This evolution allows MLOps to better meet the needs of machine learning projects and enable efficient development, deployment, and management of models.

*2.2. Traditional system integration deployment*

The traditional concept of system integration refers to the integration of computer systems, including the integration of computer hardware platform, network system, system software, tool software and application software, and the corresponding consultation, service and technical support around these systems. It is based on computer related technology reserves, with reliable products as tools, to achieve a specific combination of computer system functions of the engineering behavior. The content of system integration includes integration of technology environment, integration of data environment and integration of application program. [8]The system integration of network information system is to use advanced computer and communication technology to integrate the small operating environment supporting each information island into a large operating environment.

From the point of view of system integration, a typical network information system consists of different systems. These systems typically come from multiple vendors, include multiple incompatible hardware and software platforms, and run a variety of business, scientific, and engineering applications.

In order to connect these heterogeneous systems and port applications from one system to another, existing proprietary systems must adapt to standard interfaces and transition to open systems. What users want is interoperability between multi-vendor platforms. Therefore, the work of system integration is very important in the construction of information system projects. It integrates all kinds of resources organically and efficiently through hardware platform, network communication platform, database platform, tool platform and application software platform to form a complete workbench. However, the quality of system integration has a great impact on system development and maintenance. The basic principles to be followed in technology include: openness, structure, advancement and mainstreaming.

*2.3. Machine learning and system automation model training*

MLOps is an ML engineering culture and practice that seeks to unify ML System development (Dev) and ML System Operations [9](Ops). Practicing MLOps means advocating automation and monitoring in all steps of ML system building, including integration, testing, release, deployment, and infrastructure management.

Given relevant training data for use cases, data scientists can implement and train ML models with predictive performance on off-line retention datasets. However, the real challenge is not to build a machine learning model, but to build an integrated machine learning system and run it consistently in production. With Google's long history of producing ML services, you understand that there can be many pitfalls in operating ML-based systems in production.

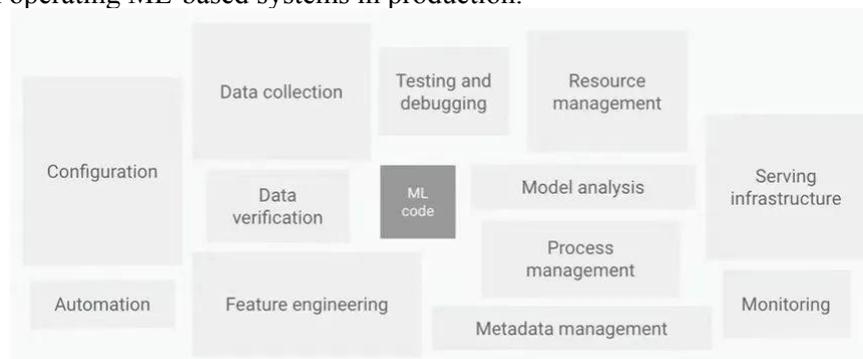

**Figure 1.** Deployment of machine learning in system automation

In Figure 1, the rest of the system consists of configuration, automation, data collection, data validation, testing and debugging, resource management, model analysis, process and metadata management, service infrastructure, and monitoring.

To develop and operate such complex systems, DevOps principles can be applied to ML systems (MLOps). This document covers concepts to consider when setting up an MLOps environment for data science practices, such as CI, CD, and CT in ML.

*2.4. Automation benefits of machine learning systems*

Machine learning systems differ from other software systems in the following ways:

Team skills: In machine learning projects, teams typically include data scientists or machine learning researchers who focus on exploratory data analysis, model development, and experimentation. These members may not be experienced software engineers who can build production-grade services.

- Development: Machine learning is experimental in nature. You should try different features, algorithms, modeling techniques, and parameter configurations to find the best fit for the problem as soon as possible. The challenge is keeping track of what works and what doesn't, and maintaining repeatability while maximizing code reusability. Testing: Testing machine learning systems is more complex than testing other software systems. In addition to typical unit and integration testing, you need data validation, trained model quality assessment, and model validation.

Deployment[10]: In an ML system, deployment is not as simple as deploying an offline trained ML model as a predictive service. ML systems may require you to deploy multi-step pipelines to automatically retrain and deploy models. This pipeline adds complexity and requires you to automate the steps that data scientists do manually before deployment to train and validate new models.

Production: Performance degradation of machine learning models is not only due to suboptimal coding, but also due to constantly changing data profiles. In other words, models can degrade in more ways than traditional software systems, and you need to take that degradation into account. Therefore, you need to keep track of aggregate statistics for your data and monitor the online performance of your model to send notifications or rollback when values deviate from your expectations.

In the past development, MLOps, as an evolution of machine learning system development and operation, introduced the principles of automation and continuous integration deployment from the basis of DevOps, and provided new methods and tools for the development, testing and deployment of machine learning models. Through continuous monitoring and feedback mechanisms, MLOps emphasizes constant attention to model performance and data changes to optimize model stability and accuracy. However, the lack of software engineering experience of machine learning team members, as well as the complexity of model deployment and sensitivity to data changes, also pose certain challenges to the practice of MLOps that require further research and resolution.

**3. Methodology**

*3.1. Automation in Model Training*

Automating the model training process is essential to simplify development and ensure repeatability. First, automation can dramatically reduce human intervention, improve efficiency, and save time and resources. By using automated tools and techniques, data scientists and machine learning engineers can offload most repetitive tasks to computers, such as data preprocessing, feature engineering, and model tuning. In this way, they can devote more energy to the design and optimization of the model, which accelerates the development cycle and improves the quality and performance of the model. Second, automation also ensures repeatability and consistency in model training. By using version control systems and automated workflows, teams can track and manage all changes during model training, including data sets, parameter Settings, and algorithm selection. This not only facilitates collaboration and communication among team members, but also ensures consistent performance of the model across different environments, improving model reliability and maintainability.

In real-world deployments, large neural networks (DNNS) face challenges due to their huge demands on resources. [11]Traditional DNNS may face hardware limitations, insufficient computing resources, and latency in actual deployments, especially in environments where edge devices or resources are limited. To overcome these challenges, compressed neural networks have become one of the key solutions. By reducing the number of parameters and computational complexity of the model, compression technology can significantly reduce the storage and computational requirements of the model, making it more suitable for real-world deployment. When compressing neural networks, it is necessary to reduce the size and complexity of the model with as little performance loss as possible. This can be achieved by pruning, quantization, low rank approximation and other techniques. At the same time, hardware optimization techniques such as hardware accelerators and dedicated processors can be utilized to further improve the efficiency and performance of the model. Therefore, the compressed neural network can not only reduce the deployment cost and delay, but also expand the application range of the model in different devices and scenarios, and promote the process of DNN productization.

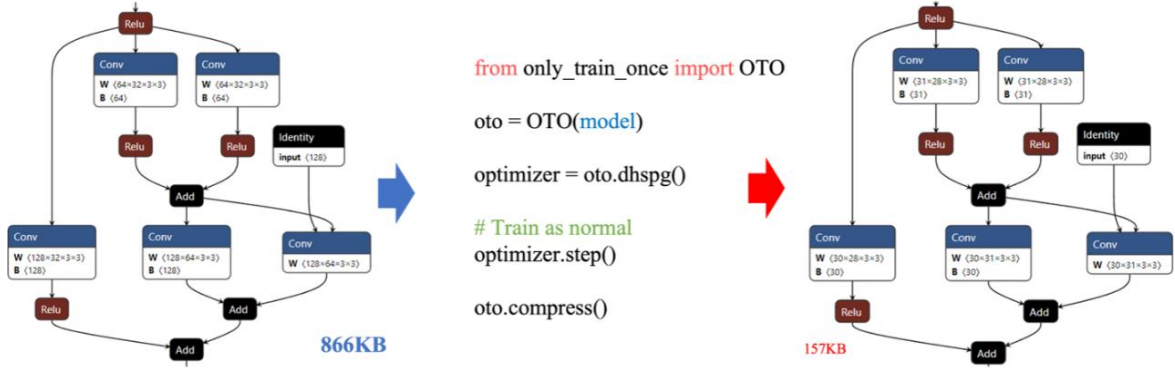

**Figure 2.** OTOv2 framework based on DNN automated deployment

Pruning is one of the most common DNN compression methods, which aims to reduce redundant structures, slim down DNN models while maintaining model performance. However, existing pruning methods often point to specific models, specific tasks, and require AI engineers to invest a lot of engineering and time effort to apply these methods to their own tasks. Therefore, to address these issues, the Microsoft team proposed the [12]OTOv2 framework, which was published in ICLR 2023. OTOv2 is the industry's first automated, one-stop, user-friendly, and versatile framework for neural network training and structural compression.

In actual deployment, large neural networks (DNNs) face the challenge of huge resource requirements. Traditional DNNs can suffer from hardware limitations, insufficient computing resources, and latency, especially in edge devices or resource-constrained environments. To overcome these challenges, compressed neural networks have become one of the solutions. By reducing the number of parameters and computational complexity of the model, compression technology can significantly reduce the storage and computational requirements of the model, making it more suitable for real-world deployment. When compressing neural networks, it is necessary to reduce the size and complexity of the model with as little loss of performance as possible. This can be achieved by pruning, quantization, low rank approximation and other techniques. At the same time, hardware optimization techniques such as hardware accelerators and dedicated processors can be utilized to further improve the efficiency and performance of the model. Therefore, the compressed neural network can not only reduce the deployment cost and delay, but also expand the application range of the model in different devices and scenarios, and promote the process of DNN productization.

In addressing these challenges, version control systems play a key role in ensuring transparency and repeatability in model training. By using a version control system, the team can track and manage all changes during model training, including data sets, parameter Settings, and algorithm selection. This tracking and management mechanism not only facilitates collaboration and communication among team members, but also ensures the consistent performance of the model in different environments, improving the reliability and maintainability of the model. Therefore, version control systems play a crucial role in the automated model training process and are a key component in ensuring high-quality, reliable model development.

### 3.2. Integration with CI/CD Pipelines

DevSecOps is a cultural approach in which every team and person working on an application considers security throughout its life cycle. It ensures that security is implemented at every stage of the application software development life cycle [13](SDLC) by embedding the required security checks into CI/CD automation using appropriate tools. But in actual deployments, with vulnerabilities emerging faster than ever before, integrating Dynamic Application security testing (DAST) into continuous integration/continuous deployment [14](CI/CD) pipelines is a game-changer, helping you consider security at an early stage, find and address security vulnerabilities as early as possible. Rather than wait until they seriously affect users before taking action.

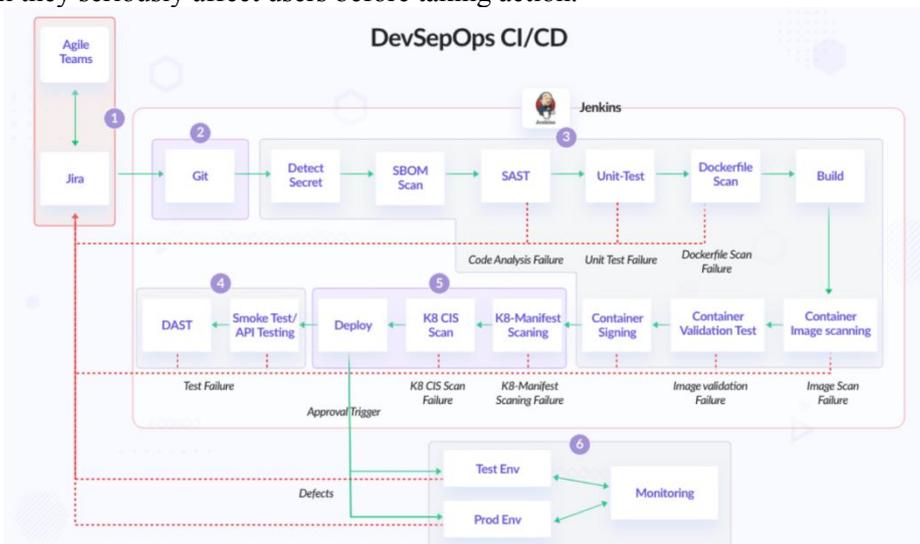

**Figure 3.** DevSecOps CI/CD pipeline architecture

CI/CD is a way to frequently deliver applications to customers by introducing automation in the application development phase. CI/CD's core concepts are continuous integration, continuous delivery, and continuous deployment.

Specifically (Figure 3), CI/CD enables continuous automation and continuous monitoring throughout the entire life cycle of the application, from the integration and testing phase to delivery and deployment. These associated transactions are often collectively referred to as "CI/CD pipelines "and are supported by development and operations teams working together in an agile manner.

In order to maximize the benefits of DAST integration in the automated deployment of machine learning systems, it can be incorporated early in the development process, such as when code reviews or new functionality is being developed. By starting early, developers can address security vulnerabilities immediately, rather than putting off fixes until later in testing. While traditional CI/CD processes are typically geared toward traditional software development, machine learning components involve complex steps such as data preprocessing, model training, and evaluation, unlike traditional software development. Therefore, integrating machine learning workflows into the CI/CD pipeline requires addressing a number of technical and methodological challenges. One of these is the management of versioned environments, as machine learning models are highly dependent on the software environment and need to ensure that the same results can be reproduced in different environments. Another challenge is containerizing machine learning components to ensure their portability and consistency across different environments. By using containerization, machine learning workflows can be packaged into containers and deployed in different environments, simplifying the process of deployment and management, and improving portability and reliability. Therefore, versioning environments and containerization are two of the main solutions to integrate machine learning components into the traditional CI/CD pipeline, which can help teams manage and deploy machine learning workflows more efficiently, accelerating the speed and quality of software delivery.

*3.3. Model Deployment and Monitoring*

In a large cluster, even routine operations can become variable, including operating system upgrades, security patch application, software package management, and custom configuration of kubelet or containerd. To ensure that all nodes in a cluster can be securely and stably updated to a consistent state, you must have not only the capability of large-scale node changes, but also the capability of auditing and rolling back change operations.

During O&M[15] operations, if the node status is inconsistent due to errors, that is, the configurations of some nodes are inconsistent with the expectations, or even multiple versions of nodes exist at the same time, the next O&M operation will fail and unexpected behaviors of the same service copies may occur on some nodes, resulting in service stability risks.

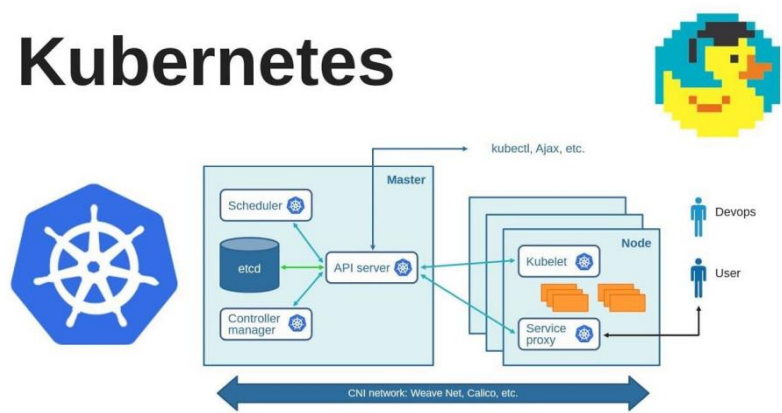

**Figure 4.** Kubernetes (K8s) automated deployment environment architecture

Large AI models are typically deployed on cloud-native environments such as Kubernetes (K8s)figure 4. With the popularity and maturity of cloud-native technologies, more and more enterprises and developers are choosing Kubernetes as the infrastructure to deploy and manage AI large model services based on the following points:

1. Resource management and Scheduling: [16]Kubernetes provides flexible and efficient resource management and dynamic scheduling capabilities, which are critical for large-scale machine learning models that rely on high-performance computing resources, especially Gpus.

2. Elastic scaling: According to the changes in the demand of the model service, K8s can automatically expand and shrink the number of Pods, so as to better use resources and ensure the high availability of services.

3. Containerized deployment: Packaging the model and its runtime environment through container technologies such as Docker, so that the model can be quickly and consistently deployed in any Kubernetes-enabled cluster.

4. Service orchestration: [17]K8s provides a complete set of service discovery, load balancing, and service governance mechanisms to help build complex microservice architectures, especially for AI applications where multiple services may be required to work together.

5. Model version management and update: With K8s rolling update, canary release and other functions, model versions can be smoothly upgraded or rolled back to reduce operation and maintenance risks.

6. Storage and persistence: K8s supports multiple types of persistent storage solutions to facilitate access and backup of large model data.

7. Monitoring and automated operation and maintenance: With various monitoring tools, performance monitoring, fault detection and self-healing can be realized during the operation of AI large models, improving the stability and operation and maintenance efficiency of the system.

Cloud-native architectures and Kubernetes have become one of the standard choices for large-scale AI deployment and management, especially when dealing with large-scale, high-performance demanding scenarios. From model training to inference services, K8s provides powerful support.

In addition, continuous monitoring and feedback loops are critical in maintaining model performance and reliability after deployment. Once a machine learning model is deployed into a production environment, its performance and behavior, as well as its interaction with the environment and data, need to be continuously monitored. This continuous monitoring can help identify potential problems and anomalies, intervene and fix them in a timely manner, and ensure that the model continues to operate stably. By monitoring model indicators and behaviors, such as model accuracy, delay, throughput, etc., model performance degradation or abnormal behavior can be detected in time, and corresponding measures can be taken, such as retraining the model, adjusting model parameters, or updating the data set. In addition, continuous monitoring can also help identify model drift and degradation phenomena, and adjust the model or data in time to adapt to changes in the environment and needs. By creating a feedback loop, you can continuously improve the model and deployment process, improve the performance and reliability of the model, and meet changing business needs and user expectations. Therefore, continuous monitoring and feedback loops are critical steps to ensure that machine learning models are running reliably in production environments, and are critical to improving the effectiveness and value of the models.

## 4. Case Studies and Best Practices

At Netflix, there are hundreds of thousands of workflows and millions of jobs running on multiple layers of the big data platform every day. Given the broad scope and intricate complexity inherent in such distributed large-scale systems, diagnosing and fixing job failures can create a considerable operational burden, even if the failed jobs represent only a small percentage of the total workload. To handle errors efficiently, Netflix has had great success using machine learning to improve the user experience, recommend content, optimize video coding, and make content distribution more efficient. Their MLOps practice is one of the keys to their success.

First, the classification of Netflix's rules-based classifier, which uses machine learning services to predict retry success probability and retry cost, and select the best candidate configuration as a recommendation; And automatically apply suggested configuration services. Its main advantages are as follows:

(a) Integrated intelligence. Instead of completely deprecating the current rules-based classifiers, autofix integrates the classifier with ML services so that it can take advantage of the best of both worlds: a rules-based classifier provides a static, deterministic classification result for each error category, based on the background of a domain expert; ML services leverage the power of ML to provide performance and cost-aware recommendations for each job. By integrating intelligence, we can well meet the requirements of fixing different errors.

(b) Full automation. The process of misclassifying, obtaining recommendations, and applying recommendations is fully automated. It provides recommendations along with retry decisions to the scheduler, and specifically uses an online configuration service to store and apply recommended configurations. In this way, no human intervention is required during the repair process.

(c) Multi-objective optimization. Automatic repair generates recommendations by considering performance (that is, the probability of retry success) and calculating cost efficiency (that is, the monetary cost of running a job) to avoid blindly recommending a configuration that consumes too many resources. For example, for memory configuration errors, it searches for multiple parameters related to memory usage for job execution and recommends a combination of linear combinations that minimize the probability of failure and computational costs.

These benefits have been validated through production deployments that fix Spark job failures. Our observations show that by applying recommended memory configurations online, automatic repair can successfully fix about 56% of memory configuration errors without human intervention; At the same time, it reduces costs by about 50% because it can recommend new configurations for successful memory configuration and prevents unnecessary retries for uncategorized errors. We also note great potential for further improvements through model tweaks.

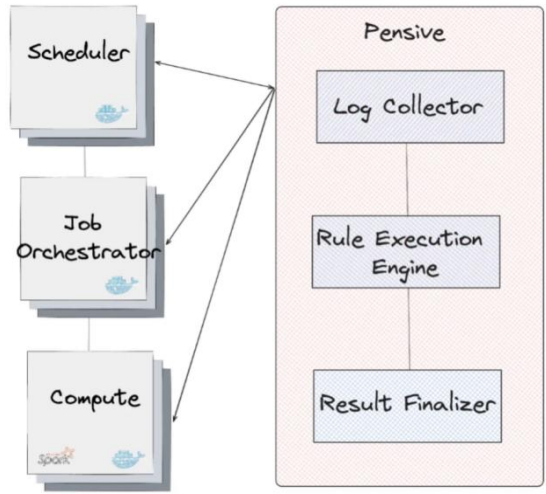

Figure 5. Pensive in Netflix Data Platform

Figure 5 shows the misclassification service, Pitive, in the Netflix data platform. It utilizes rule-based classifiers and consists of three components:

The log collector is responsible for extracting logs from different platform layers, such as schedulers, job choreographers, and compute clusters, for error classification. (Rule priority is determined by rule ordering. The first rule has the highest priority.) On the other hand, if there is no rule match, the error is considered unclassified.

*4.1. Netflix has accumulated many best practices and lessons learned in the practice of MLOps.*
**These include:**

1. Automation and Continuous integration: Netflix emphasizes automation and continuous integration, leveraging the CI/CD pipeline to automate model training, evaluation, and deployment. This automated process increases efficiency, reduces human error, and ensures rapid iteration and updating of models.

2. Containerized deployment: Netflix containerizes models and applications and leverages Kubernetes for deployment and management. Through containerization, they are able to achieve rapid deployment, elastic scaling, and high availability of models, while ensuring consistency and portability of the environment.

3. Real-time monitoring and feedback: Netflix has established a real-time monitoring and feedback mechanism to detect and resolve problems in a timely manner by monitoring model performance, user feedback, and system logs. This continuous monitoring and feedback loop helps to improve model stability and reliability, and to adjust models and services in a timely manner.

4. Refined experiment management: Netflix values experiment management and version control to ensure that each model has clear traceability and repeatability. They utilize advanced experiment management tools and processes to manage model versions, parameters, and results for effective comparison and selection.

Through these best practices and lessons learned, Netflix has not only overcome common challenges in MLOps, but also improved the efficiency and quality of workflows, laying a strong foundation for innovation and success. The successful application of these strategies provides a valuable reference for other organizations to build and optimize their own MLOps practices.

## 5. Conclusion

The content of the article reveals the challenges faced by machine learning in practical applications and introduces the importance of MLOps as a solution. By automating model training and deployment and integrating into traditional CI/CD pipelines, the complexity and challenges of deploying machine learning models in production environments can be effectively addressed. In addition, the paper emphasizes the importance of continuous monitoring and feedback loops in maintaining model performance and reliability. These methods and tools provide effective solutions for the development, deployment, and management of machine learning models, thus accelerating the model development cycle and improving the quality and performance of the models.

Looking ahead, as machine learning technologies continue to evolve and the range of applications expands, we can expect more innovation and progress. Machine learning not only plays an important role in improving business efficiency, promoting innovation in the advertising industry, and improving medical technology, but also brings great potential and benefits to human society. By applying machine learning technology more widely, we can enable smarter and more efficient decisions and services, contributing to the sustainable development of society. The development of artificial intelligence will bring more convenience and well-being to mankind, and we should continue to be committed to promoting the innovation of machine learning technology to better meet human needs and achieve social progress and development.